\documentclass[%
 aip,
amsmath,amssymb,
 reprint,%
]{revtex4-1}

\usepackage{graphicx}% Include figure files
\usepackage{dcolumn}% Align table columns on decimal point
\usepackage{bm}% bold math
\usepackage[utf8]{inputenc}
\usepackage[T1]{fontenc}
\usepackage{mathptmx}
\usepackage{etoolbox}
\usepackage{gensymb}

\makeatletter
\def\@email#1#2{%
 \endgroup
 \patchcmd{\titleblock@produce}
  {\frontmatter@RRAPformat}
  {\frontmatter@RRAPformat{\produce@RRAP{*#1\href{mailto:#2}{#2}}}\frontmatter@RRAPformat}
  {}{}
}%
\makeatother
\begin{document}

\preprint{AIP/123-QED}

\title[]{Strong coupling between quantized magnon modes in a YIG microstructure and microwaves in a superconducting resonator}
% Force line breaks with \\

\author{Seth W. Kurfman}
\affiliation{Institut für Physik, Martin Luther Universität Halle-Wittenberg, Halle, Germany 06120}

\author{Philipp Geyer}%
\affiliation{Institut für Physik, Martin Luther Universität Halle-Wittenberg, Halle, Germany 06120}

\author{Anoop Kamalasanan}%
\affiliation{Institut für Physik, Martin Luther Universität Halle-Wittenberg, Halle, Germany 06120}

\author{Karl Heimrich}%
\affiliation{Institut für Physik, Martin Luther Universität Halle-Wittenberg, Halle, Germany 06120}

\author{Kwangyul Hu}%
\affiliation{Department of Physics and Astronomy, University of Iowa, Iowa City, Iowa, USA, 52242}

\author{Tharnier O. Puel}%
\affiliation{Department of Physics and Astronomy, University of Iowa, Iowa City, Iowa, USA, 52242}

\author{Frank Heyroth}%
\affiliation{Interdiszinplinäres Zentrum für Materialwissenschaften, Martin Luther Universität Halle-Wittenberg, Halle, Germany 06120}

\author{Michael E. Flatté}%
\affiliation{Department of Physics and Astronomy, University of Iowa, Iowa City, Iowa, USA, 52242}
\affiliation{Department of Applied Physics and Science Education, Eindhoven University of Technology, Eindhoven, The Netherlands, 5612 AP}

\author{Georg Schmidt}%
\affiliation{Institut für Physik, Martin Luther Universität Halle-Wittenberg, Halle, Germany 06120}
\affiliation{Interdiszinplinäres Zentrum für Materialwissenschaften, Martin Luther Universität Halle-Wittenberg, Halle, Germany 06120}
\affiliation{Halle-Berlin-Regensburg Cluster of Excellence CCE, Germany}
\email{georg.schmidt@physik.uni-halle.edu.}

\date{\today}% It is always \today, today,
%TC:ignore
\begin{abstract}
Strong-coupling experiments based on magnons enable the exploration into on-chip demonstrations involving numerous long-lived excitations. Yttrium iron garnet (YIG) has been considered for decades as a gold standard material for magnonics due to its low-loss magnonic properties. While YIG has successfully demonstrated strong-coupling in macroscopic device geometries, the strong coupling of magnons in truly sub-10 micron YIG structures to date has not yet been realized. This obstacle is due to the difficulty producing large enough effective magnonic mode volume necessary primarily due to thickness limitations of YIG deposition and device fabrication techniques. Here, we demonstrate the use of a microplatelet of YIG, manufactured from a single crystal of YIG via focused ion beam (FIB) techniques, placed on a constricted inductive line of an optimized superconducting lumped element LC resonator to achieve strong coupling between numerous magnon modes and the LC resonator photons. These experimental findings are qualitatively backed by micromagnetic simulations and quantitatively supported by analytical calculations to identify the magnon modes corresponding to the experimentally observed anti-crossings in the microwave transmission signal. Further, we show that these anti-crossings remain even at incredibly low device input powers ($\leq 10$  fW). The fabrication techniques and device geometry enable the deterministic use of numerous confined magnon modes in micron-scale YIG structures for various magnetic field strengths and orientations at substantially reduced device powers. The results here establish a foundational path forward to achieving efficient magnon-based strong-coupling experiments in micron-scale YIG magnetic elements for effective on-chip studies. 
\end{abstract}
%TC:endignore
\maketitle

%\section{Introduction}\label{sec1}

\begin{figure*}
    %\centering
    \includegraphics[width=\linewidth]{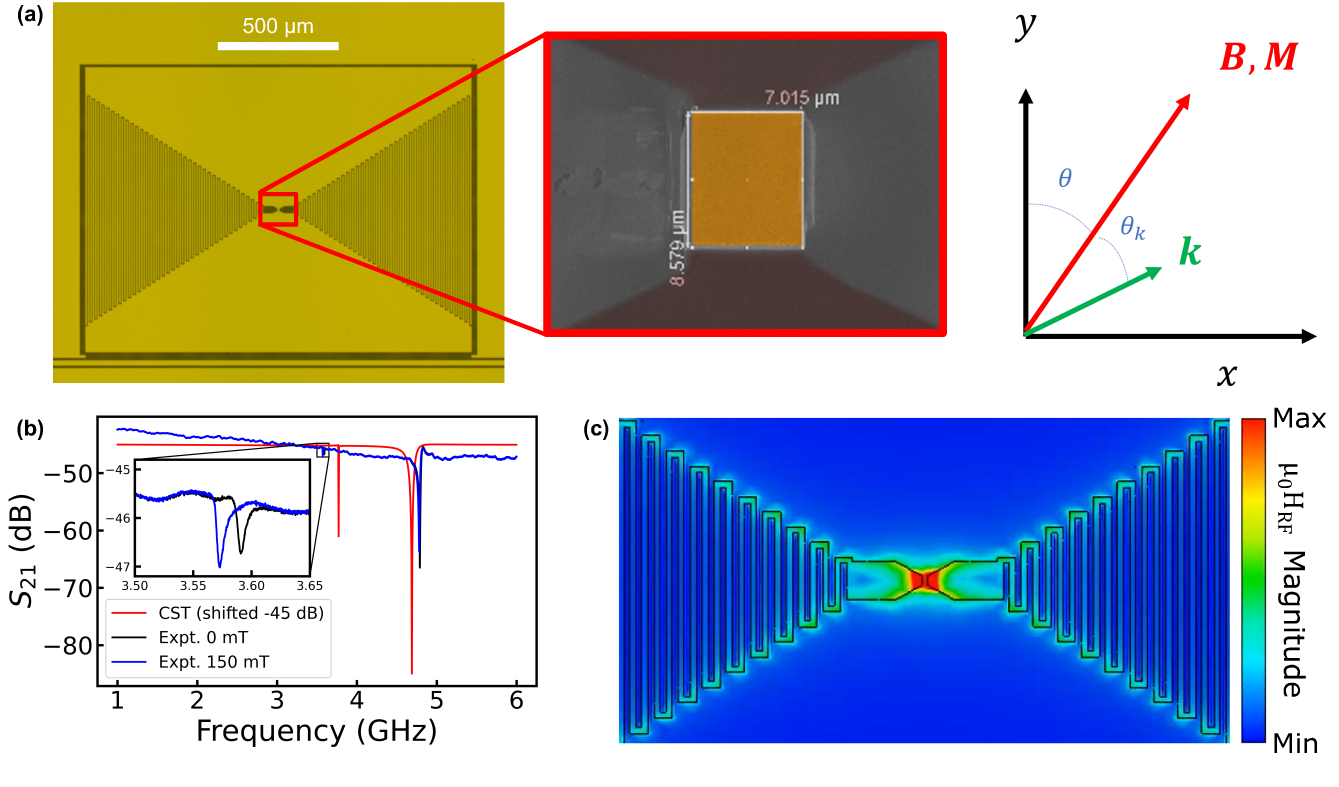}
    \caption{Overview of the fabricated device and basic characterization. (a) Optical image of the resonator device and false-colored SEM image of the YIG platelet on top of the inductive line of the resonator. The defined coordinate system on the right shows the angle of the magnetic field $\mathbf{B}$ and magnetization $\textbf{M}$ direction $\theta$ with respect to the axis of the inductive line, and $\theta_k$ indicates the angle between the magnetization and the spin wave $k$-vector $\mathbf{k}$. The applied microwave field $\mu_0 H_{RF}$ is always applied along the $\pm x$-axis. (b) CST simulations of the bow-tie resonator for the inductive mode and comparison to experimental data. Simulated $S_{21}$ transmission through the signal line of the simulated DUT (red and shifted by -45 dB for clarity) compared to experimental VNA $S_{21}$ scans of the DUT at $T= 1.4$ K for 0 mT (black) and 150 mT (blue) applied magnetic field. %Note the dip in the CST simulations corresponding to the mode used for the experiments is much deeper ($\sim$9 dB) than the experimental signature which will be considered later in the power dependence and non-linear driving discussion.
    The slight frequency discrepancy of the peaks is simply due to slight deviations of the actual device during fabrication compared to the simulations. (c) Heatmap from CST simulations showing the strength of the RF magnetic field around the constriction. The red color indicates a photon mode volume highly confined to the region around the constriction. A more complete comparison of the inductive and capacitive modes can be found in the Supplementary Material.}
    \label{fig:fig1_SampleOverviewAndExperimentSpectraVsCST}
\end{figure*}

The low-lying excitations in magnetically ordered materials known as spin waves (or their quanta, magnons) enable investigations into numerous fundamental interactions with spin, charge, and vibrational degrees of freedom, and demonstrate potential for low-power and high-frequency information processing \cite{Awschalom_2021_IEEETransOnQE_QuantumEngineeringWithHybridMagnonicSystemsAndMaterials,Chumak_2022_IEEEtransOnMagn_AdvancesInMagneticsRoadmapOnSWcomputing, Barman_2021_JPhysCondMat_The2021MagnonicsRoadmap, Flebus_2023_JAP_RecentAdvancesInMagnonics, Pirro_2021_NatureReviews_AdvancesInCoherentMagnonics, Kruglyak_2010_JPhysD_Magnonics, Chumak_2017_JPhysDApplPhys_MagnonicsCrystalsForDataProcessing}.  Magnon-based strong coupling experiments, wherein information is coherently exchanged between magnons and other resonant excitations, are one such group of fundamental studies with application potential in, for example, quantum information science and engineering (QISE) systems \cite{Lachance-Quirion_2019_ApplPhysExpress_HybridQuantumSystemsBasedOnMagnonics}. Such studies have already been theoretically predicted \cite{Soykal2010, Candido2020} and experimentally demonstrated strong coupling of magnons with on-chip LC resonator photons \cite{Huebl_2013_PRL_HighCooperativityInCoupledMicrowaveResonatorFerrimagneticInsulatorHybrids, Li2019, Hou2019, Baity2021, Xu_2024_AdvScience_StrongPhotonMagnonCouplingUsingALithographicallyDefinedOrganicFerrimagnet}, cavity microwave photons \cite{Zhang_2014_PRL_StronglyCouplingMagnonsAndCavityMicrowavePhotons,Tabuchi_2014_PRL_HybridizingFerromagneticMagnonsAndMicrowavePhotonsInTheQuantumLimit}, optical photons \cite{Zhu_2020_Optica_WaveguideCavityOptomagnonicsForMicrowaveToopticsConversion}, phonons \cite{Zhang_2016_SciAdv_CavityMagnomechanics,Zhang_2016_PRL_OptomagnonicWhisperingGalleryMicroresonators}, superconducting qubits \cite{Lachance-Quirion_2020_Science_EntanglementBasedSingleShotDetectionOfASingleMagnonWithASCqubit}, and other magnon modes \cite{Chilcote_2019_APLMater_SpinWaveConfinementAndCouplingInOrganicBasedMagneticHeterostructures}. Achieving strong coupling where information is coherently exchanged between coupled resonators requires long-lived (low loss) excitations and a strong interaction between the resonators, wherein the decay rates of the coupled modes must be less than the coupling rate between them \cite{Zhang_2014_PRL_StronglyCouplingMagnonsAndCavityMicrowavePhotons}. The coupling strength is enhanced when the (volumetric) mode overlap of the magnon and coupled resonator modes is increased \cite{Zhang_2014_PRL_StronglyCouplingMagnonsAndCavityMicrowavePhotons}. Currently, however, these magnon-based strong coupling experiments are predominantly limited to macroscopic (mm- and cm-scale) devices with the smallest magnetic elements extending laterally into the range of tens or hundreds of $\mu m$ (for example, Refs. \onlinecite{Li2019,Hou2019, Baity2021}). This is primarily due to constraints in fabricating truly micron-scale magnetic structures with low-damping and sufficient magnetic volume, as well as resonator systems with confined mode volumes to enhance mode overlap. Although these geometric length scales provide a sufficient number of spins for the coupling, which scales as $\sqrt{N}$ with $N$ being the number of spins in the magnetic system, a clear drawback is the need to miniaturize and couple multiple elements in a practical chip design. Consequently, extending such magnon-based studies towards on-chip device infrastructures with practical application uses requires reducing the lateral size of both the magnetic element and the coupled resonator structure.

The canonical material for ultra low-loss magnonic studies is yttrium iron garnet (YIG), known for its low-loss (high-$Q$) ferromagnetic resonance (FMR) in the GHz regime \cite{Serga_2010_JPhysD_YIGmagnonics, Schmidt_2020_PhysStatSolB_UltraThinFilmsOfYIGwithVeryLowDamping-AReview, Chumak_2022_IEEEtransOnMagn_AdvancesInMagneticsRoadmapOnSWcomputing}. Despite its remarkable magnonic properties and well-established use in hybrid quantum magnonic systems \cite{Lachance-Quirion_2019_ApplPhysExpress_HybridQuantumSystemsBasedOnMagnonics}, several hurdles remain when utilizing YIG in devices, particularly the complexity of integrating it with conventional device infrastructures and the challenge of maintaining sufficient magnetic volume (i.e., number of spins) as lateral dimensions are reduced for miniaturization.  Limitations in producing thick YIG films by off-axis sputtering or pulsed laser deposition (PLD) accordingly make the volume requirements challenging when reducing the lateral dimensions. While thick films of YIG can be produced via liquid phase epitaxy (LPE), lithographic techniques needed to produce microstructures introduce roughness to the substrate and therefore negatively impact the quality of on-chip superconducting resonators. Furthermore, the GGG substrate paramagnetic spins and stray fields at low-temperatures render superconducting materials ineffective, therefore requiring the use of alternative substrate materials \cite{Sparks_Physrev_1964_Ferromagnetic_RelaxationTheory, Seiden_PhysRev_1964_FerrimagneticResonanceRelaxationInRareEarthIronGarnets,Serha_2024_npjSpintronics_MagneticAnisotropyAndGGGsubstrateStrayFieldInYIGfilomsDownTomKtemps, Knauer_2023_JAP_PropagatingSpinWaveSpectroscopyInALPEnmThickYIGfilmAtmKTemps,Serha_2026_NatComm_YSGAG-TheIdealSubstrateForYIGinQuantumMagnonics,Guo_2023_NanoLett_StrongOnChipMicrowavePhotonMagnonCouplingUsingUltralowDampingEpitaxialYIGfilmsAt2K}. Accordingly, the limitations discussed above highlight a unique challenge in the direct integration of high-quality, (sub-)micron YIG structures that can achieve strong coupling with on-chip superconducting microwave resonators.

Here, we integrate a YIG platelet microstructure ($7$ $\mathrm{\mu}$m $\times$ $8.6$ $\mathrm{\mu}$m $\times$ $790$ nm) with an optimized on-chip superconducting lumped element LC resonator as seen in Fig. \ref{fig:fig1_SampleOverviewAndExperimentSpectraVsCST}(a). The dimensions of the YIG platelet are chosen such that the resonance conditions of the confined magnon modes with discrete wavenumbers (wavelengths) become resolvable and clearly separated. Using focused ion beam (FIB) milling techniques (see Supplementary Material), a YIG microstructure is fabricated from a high-quality LPE YIG film. This platelet is placed directly on a constricted inductive line of an optimized superconducting LC resonator structure, where the constriction (width $w = 4$ $\mathrm{\mu}$m) locally increases the RF magnetic field. 
At the same time the RF field is concentrated to a smaller volume enhancing  the effective coupling strength between the resonator and magnon modes in the small YIG platelet, allowing for excitation power levels as low as $10\,fW$ (see Supplementary Material). The fact that the inductive line is smaller than the platelet reduces the magnon-photon mode overlap compared to the optimal case; however, it allows for the $z$-component of the RF-field at the edge of the inductive line to strongly contribute to the excitation of antisymmetric modes as we show below.

\begin{figure*}
    \centering
    \includegraphics[width = \linewidth]{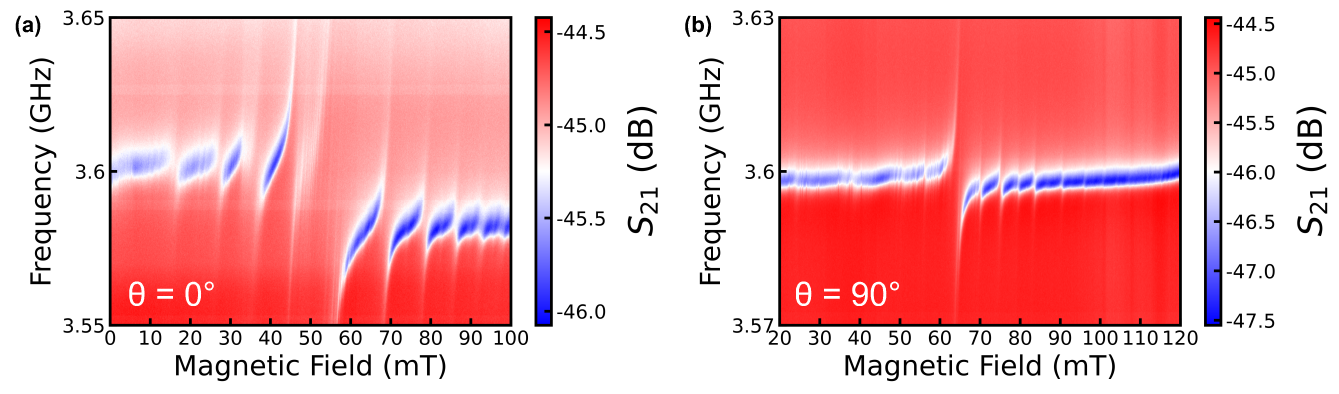}
\caption{VNA $S_{21}$ frequency-sweep measurements as a function of magnetic field. Panels (a) and (b) show $S_{21}$ microwave transmission as a function of applied external magnetic field $\mu_0 H_{ext}$ for (b) $\theta = 0\degree$ and (c) $\theta = 90\degree$. The $3.6$ GHz mode of the LC resonator shows numerous avoided crossings, associated with the Kittel mode ($\mu_0 H_{ext}\sim 50$ mT) and various confined modes of the YIG platelet. The baseline of the $S_{21}$ data  $\sim$ -44.5 dB is a result of the attenuation by two -20 dB attenuators and the cables.}
    \label{fig:fig2_Expt-Spectra}
\end{figure*}

Microwave simulations (CST Studio Suite) of the superconducting LC resonator are performed using the drawing files from the lithography process and creating realistic conditions for the device including the substrate and copper sample mounting box. The $S_{21}$ signal of the simulated device reveal two distinct modes which align well with the experimental $S_{21}$ signal measured by a Vector Network Analyzer (VNA) where the dips correspond to resonant modes in the resonator, as seen in Fig. \ref{fig:fig1_SampleOverviewAndExperimentSpectraVsCST}(b). The primary mode of interest in the device at $f \sim 3.6$ GHz is identified and selected for use in these measurements as the RF magnetic field (i.e. photon mode volume) is spatially reduced and primarily isolated to the inductive element line as shown in Fig. \ref{fig:fig1_SampleOverviewAndExperimentSpectraVsCST}(c). In the experiment, the resonator shows a loaded $Q$-factor of 1230 corresponding to a linewidth of 2.92 MHz. 

%\section{Results}
%\subsection{Measurement Results}

To investigate any coupling between the LC resonator and magnons in the YIG platelet, the VNA $S_{21}$ signal is measured for a range of applied in-plane magnetic field strengths and orientations (i.e. varying $\theta$ as described in Fig. \ref{fig:fig1_SampleOverviewAndExperimentSpectraVsCST}(a)). The resulting $S_{21}$ spectra versus the external magnetic field shown in Fig. \ref{fig:fig2_Expt-Spectra} reveal numerous anti-crossing features indicative of coupling between the LC resonator photons and magnon modes in the YIG platelet. Of particular interest in Fig. \ref{fig:fig2_Expt-Spectra}(a) is the dramatic and notably large anti-crossing and signatures between $45- 55$ mT when $\theta = 0\degree$. This gap corresponds to the coupling of the resonator to the quasi-uniform mode. In the gap, faint additional features are visible, which may be related to modes with additional quantization perpendicular to the platelet like perpendicular standing spin waves (PSSW).  
In this geometry, we observe anti-crossings at fields lower as well as higher as that of the uniform mode, reflecting the magnon dispersion with separate Damon-Eshbach and Backward Volume modes. 

Upon rotating the external, in-plane magnetic field to $\theta = 90\degree$ (see Fig. \ref{fig:fig2_Expt-Spectra}(b)), various anti-crossing signatures are also seen in the spectrum although only at higher magnetic fields than that of the quasi-uniform mode for $\theta = 0 \degree$.   

\begin{figure*}
    %\centering
    \includegraphics[width=\linewidth]{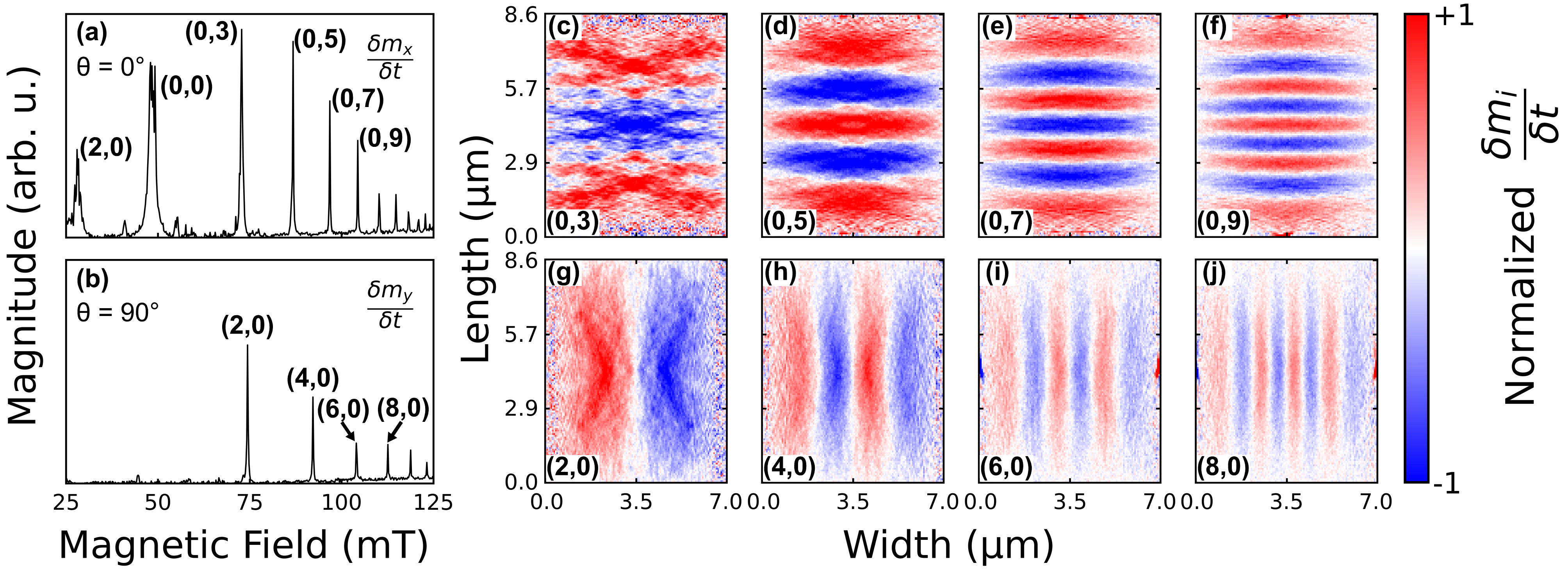}
    \caption{
    Simulated magnon spectra showing the dynamic magnetization components (a) $\delta m_x/\delta t$ for $\theta = 0\degree$ and (b) $\delta m_y/\delta t$ for $\theta = 90\degree$ calculated by micromagnetic simulations for a 7 micron width waveguide. Each peak in the spectra corresponds to a magnon mode, with specific modes labeled by their indices $(m_w,m_l)$. The panels (c-f) show mode maps for individual high field modes from (a) and (g-j) from (b) in a horizontal layer centered in the platelet. Note, the two orientations permit only odd BVMSW-like modes for $\theta = 0\degree$ and only even modes for $\theta = 90\degree$, aligning with the expectation based on the spin wave selection rules from symmetry considerations. The (2,0) peak in panel (a) corresponds to the lowest order Damon-Eshbach mode.}
    \label{fig:fig3_MuMax-Spectra-and-Mode-Maps}
\end{figure*}

In order to better understand the experimental spectra, $\mathsf{MuMax3}$ micromagnetic simulations \cite{Vansteenkiste2011,Vansteenkiste_2014_AIP-Adv_TheDesignAndVerificationOfMuMax3,Exl2014, Leliaert2017} were performed that mimic the experiment (see Supplementary Material) to produce an effective spin wave spectrum expected from the devices. By taking the difference in the respective relevant components of the magnetization between each simulation timestep, the discrete difference quotient of $\delta m_x/\delta t$ ($\theta = 0\degree$) and $\delta m_y/\delta t$ ($\theta = 90\degree$) are plotted as a function of the applied DC magnetic field as shown in Fig. \ref{fig:fig3_MuMax-Spectra-and-Mode-Maps}(a) and (b) for the  $\theta = 0\degree$ and $\theta = 90\degree$ orientations, respectively. Note that the strength of the RF field in the $\mathsf{MuMax3}$ simulations was specifically identified and chosen to align well with the expected field strength in the YIG platelet for the applied device power from the VNA as determined from the microwave simulations of the device. Further, typical material parameters for YIG at near-zero temperature are used in the simulation as outlined in the Supplement. As such, the $\mathsf{MuMax3}$ simulations should, in principle, reasonably represent the expected experimental conditions. Notably, the simulated spectra qualitatively reproduce several features that are seen in the experimental spectra. For the $\theta = 0\degree$ orientation, the spectra in Fig. \ref{fig:fig3_MuMax-Spectra-and-Mode-Maps}(a) reproduces (i) the broad main resonance with additional features and (ii) numerous resonances at fields above and below the large resonance features seen in Fig. \ref{fig:fig2_Expt-Spectra}(a). Similarly, for the $\theta = 90\degree$ orientation, the sharp peaks in the spectra of Fig. \ref{fig:fig3_MuMax-Spectra-and-Mode-Maps}(b) effectively appear only at higher fields which is qualitatively identical to what is seen in Fig. \ref{fig:fig2_Expt-Spectra}(b). While these spectra provide reasonable qualitative support of the experimental results seen in Fig. \ref{fig:fig2_Expt-Spectra}, the resonance fields deviate from the experiment. This deviation may for example result from surface and edge damage inducing local changes in saturation magnetization and for possible differences in the pinning conditions between simulation and experiment.

%\subsection{Spinwaves and Selection Rules} 

The resonance peaks in the $\mathsf{MuMax3}$ are identified by analyzing and mapping the magnetization dynamics in each cell of the simulated structure and can be characterized by two indices $(m_w,m_l)$ for quantization along the width ($w$) and length ($l$). We define $k_x=m_w\pi/w$, $k_y=m_l\pi/\ell$, so an even (odd) index $m$ corresponds to an even (odd) number of antinodes in the corresponding direction. (0,0) is the quasi uniform mode. Mode maps for the sharp,  distinct resonances at high fields are shown in Fig. \ref{fig:fig3_MuMax-Spectra-and-Mode-Maps}(c-f) for the $\theta=0\degree$ orientation and Fig. \ref{fig:fig3_MuMax-Spectra-and-Mode-Maps}(g-j) for the the $\theta=90\degree$ orientation, all indicated with their respective mode indices. 

For $\theta = 0\degree$, we observe quantized spin wave modes for magnetic fields above and below the quasi uniform mode. For higher fields the dispersion relation allows only for backward volume like modes (BVMSW-like) with $k$ in the $y$-direction. Because the RF field is homogeneous in this direction, it can only couple to modes with finite total magnetization, namely with odd $m_l$ (i.e. $m_l^{\theta = 0\degree} = \{1,3,5,...\}$).
At fields below the uniform mode, only Damon-Eshbach like (DE-like) modes with $k$ in the $x$-direction across the waveguide are allowed. In this direction, the $z$-component of the RF-field is antisymmetric which now also allows for the excitation of modes that are antisymmetric across the inductive line with even $m_w$. Indeed, in the experiment we observe several modes with $m_w=2$ and additional quantization in $y$-direction. Modes with $m_w>2$ are out of the field range for the resonance frequency.

In $\theta=90\degree$ orientation, the $x$-component of the excitation field is along the external magnetic field and cannot excite a precession. The antisymmetric $z$-component mentioned above, however, can still couple, but only to modes with finite $k_x$ (BVMSW-like) which are even in the $x$-direction ($m_w^{\theta = 90\degree} = \{2,4,6,...\}$) and thus also antisymmetric. Note that the confined uniform mode, though indexed with $m_l=m_w=0$ has the symmetry of an odd mode and is thus not excited here.   

This is of particular interest to the magnonics community for future device applications, as the lack of a uniform RF excitation in fact extends and improves the device functionality in multiple geometric orientations and multiple magnetic fields for a confined structure. In larger systems, where the modes are more closely degenerate, a uniform RF excitation might be preferred in order to remove these additional modes that effectively broaden the resonance linewidths. 

The features within the anti-crossing of the (0,0) mode may be caused by the confinement in three dimensions that allows a variety of mixed modes with simultaneous higher-order quantization along more than one axis, e.g. PSSW with additional lateral quantization. These will be described elsewhere, but an example is given in the Supplementary Material. Such in-gap features have, for example, been observed in Refs. \onlinecite{Zhang_2016_JAP_SuperstrongCouplingofThinFilmMagnetostaticWavesWithMicrowaveCavity, Xu_2024_AdvScience_StrongPhotonMagnonCouplingUsingALithographicallyDefinedOrganicFerrimagnet} and are supposed to add to the coupling strength of the main mode. As the behavior of the quantized modes for both orientations for magnetic fields at and above the quasi-uniform mode align well with the experimental results, it the reasonable to assign the various modes seen in the $\mathsf{MuMax3}$ simulations to those measured in the physical device.

%\section{Theory and Background}

To understand coupling between magnon modes and the resonator, we first consider the linearized Landau-Lifshitz equation describing the spin wave dynamics:
\begin{eqnarray}
    \dfrac{\partial \mathbf{M}}{\partial t}=-\gamma \mu_0 \mathbf{M}\times\mathbf{H}_{eff},
\end{eqnarray}
where $\mathbf{M}$ is the magnetization, $\mathbf{H}_{eff}$ is the effective magnetic field, $\gamma$ is the gyromagnetic ratio and $\mu_0$ is the magnetic permeability. 

Using the formalism developed by Kalinikos and Slavin \cite{Kalinikos1986}, the magnon mode frequencies are given by
\begin{eqnarray}    {\omega}^2=&\left(\omega_H+\alpha\omega_Mk^2+\omega_MP_{00}\sin^2\theta\right) \nonumber \\
    &\times\left(\omega_H+\alpha\omega_Mk^2+\omega_M\left(1-P_{00}\right)\right),
\end{eqnarray}
where $\omega_H=\gamma \mu_0 H_e$, $\omega_M=\gamma\mu_0 M_s$, $k=\sqrt{{k_x}^2+{k_y}^2}$, $k_x=m_w\pi/w$, $k_y=m_l\pi/\ell$, $\tan\theta=k_y/k_x$ and $P_{00}=1-\left(1-e^{-kt}\right)$ with $t$ being the thickness of the film, and $\alpha=2.96\times10^{-16}~\text{m}^2$ is the exchange constant.
Here, $\mu_0 H_e$ is the external magnetic field, $\mu_0 M_s$ is the saturation magnetization, and $m_w$ and $m_l$ are the magnon mode indices along the $x$- and $y$-directions, respectively.
The total dynamical state of the coupled system can then be expanded as
\begin{eqnarray}
    \left|\Psi\right>=a_r\left|r\right>+\sum_n a_n\left|n\right>
\end{eqnarray}
where $a_r$ is the amplitude of the resonator state $\left|r\right>$, and $a_n$ is the amplitude of $n$th magnon mode $\left|n\right>$.  The effective Hamiltonian of the coupled magnon-resonator system is written as
\begin{eqnarray}
    \mathcal{H}=\begin{pmatrix}
    \omega_r & g_0 & g_1 & g_2 & \cdots \\
    g_0 & \omega_0 & 0 & 0 & \cdots \\
    g_1 & 0 & \omega_1 & 0 & \cdots \\
    g_2 & 0 & 0 & \omega_2 & \cdots \\
    \vdots & \vdots & \vdots & \vdots & \ddots
    \end{pmatrix}
    \label{eqn:Ham}
\end{eqnarray}
where $\omega_r$ is the resonator frequency, $\omega_n$ is the frequency of the $n$th magnon mode, and $g_n=\left<r\right|V\left|n\right>$ is the coupling coefficient associated with interaction $V$.

%\subsection{Comparison of the Analytical Calculation and Experiments}

\begin{figure}
   \centering
    \includegraphics[width=\linewidth]{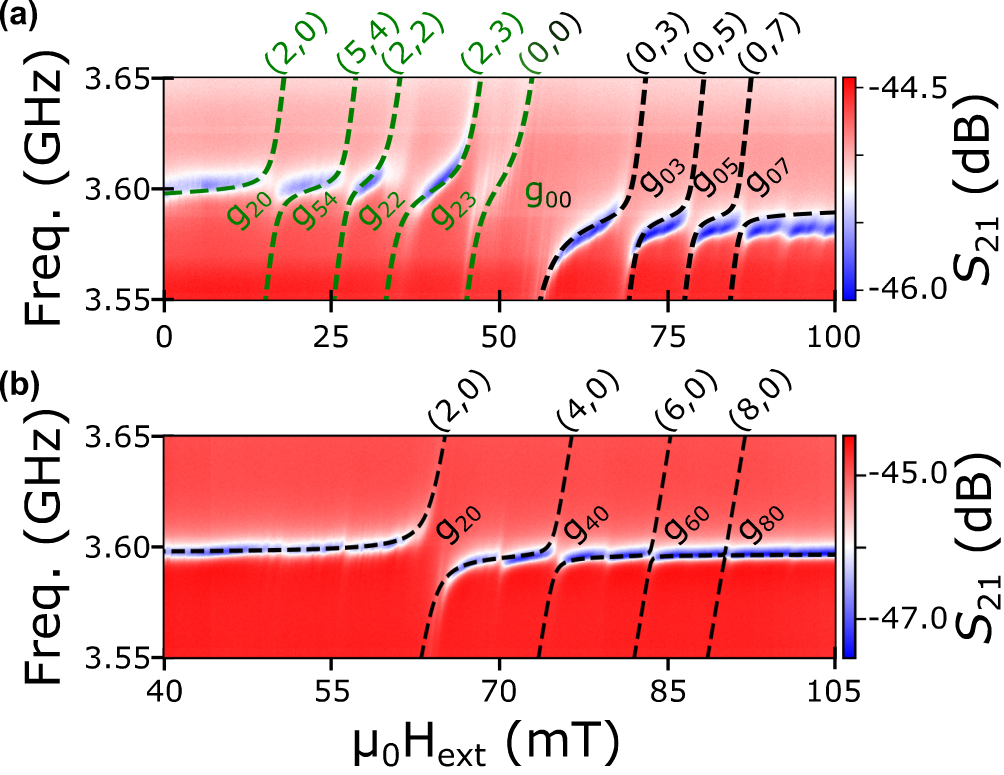}
    \caption{Measured (color plots - same as Fig. \ref{fig:fig2_Expt-Spectra}) and calculated (dashed lines) spectra of magnon modes strongly coupled to the resonator. The green dashed lines represent DE-like and hybrid modes, and the black dashed lines represent BVMSW-like modes. For each anti-crossing fit, the mode indices are indicated. (a) $\mu_0 H_{ext}$ is applied along the $y$-axis ($\theta=0^{\circ}$). (b) $\mu_0 H_{ext}$ applied is along the $x$-axis ($\theta=90^{\circ}$). Note in both cases there are signatures of weakly-coupled BVMSW-like modes between the fits in both orientations - these are likely the result of additional inhomogeneities of the RF field in the actual device due to the variation of the inductive line width in the fabricated device as shown in the Supplement.}
    \label{fig:fig4_theoryFits}
\end{figure}

By solving the eigenvalue problem of the Hamiltonian in Eq. \ref{eqn:Ham}, we obtained the spectra of the coupled magnon-resonator system depicted as dashed lines in Fig. \ref{fig:fig4_theoryFits}. The black and green dash lines correspond to BVMSW and DE-like/hybrid modes with two dimensional quantization, respectively. Figure \ref{fig:fig4_theoryFits}(a) shows the spectrum for $\mu_0 H_e\parallel\hat{y}$ corresponding to the $\theta = 0\degree$ orientation with $\mu_0 M_s=245~\text{mT}$. For the BVMSW modes, we obtained the coupling constants $g_{00}=60~\text{MHz}$, $g_{03}=20~\text{MHz}$, $g_{05}=15~\text{MHz}$ and $g_{07}=12~\text{MHz}$. For the hybrid modes, we found $g_{23}=30~\text{MHz}$, $g_{22}=30~\text{MHz}$, $g_{54}=15~\text{MHz}$ and $g_{20}=20~\text{MHz}$. Considering the number of spins in our platelet, the uniform mode exhibits a coupling per spin of $g_s \approx 61$ Hz \cite{Hou2019,Li2019,Baity2021}. Although we cannot determine the exact linewidth for the quasi uniform mode, we can make a very conservative estimate. At 3.6 GHz the material from which the platelet was fabricated typically exhibits an FMR linewidth of a few MHz. Even if we assume a value as high as 20 MHz allowing for deterioration during the FIB process, we would still be in the range of strong coupling for $g_{00}$ and at least close for some of the other anti-crossings. 

Figure \ref{fig:fig4_theoryFits} (b) illustrates the spectrum for $\mu_0 H_e\parallel\hat{x}$ corresponding to the $\theta = 90\degree$ orientation with $\mu_0 M_s=270~\text{mT}$. In this case, the extracted coupling constants are $g_{20}=25~\text{MHz}$, $g_{40}=10~\text{MHz}$, $g_{60}=3~\text{MHz}$ and $g_{80}=1~\text{MHz}$. These values are similar to those obtained for the $\theta = 0\degree$ configuration.

For the BVMSW modes, the coupling strength decreases with increasing mode order, consistent with trends reported in previous studies \cite{Zhang_2014_PRL_StronglyCouplingMagnonsAndCavityMicrowavePhotons,Zhang_2016_JAP_SuperstrongCouplingofThinFilmMagnetostaticWavesWithMicrowaveCavity}. In contrast, the hybrid modes do not exhibit a clear monotonic relationship between mode order and coupling strength, likely attributed to the mode overlap factor accounting for the mode quantization along the width and length. Further, note that all the considered modes here are for the lowest order thickness-quantized modes (PSSWs). However, these additional modes cannot be ruled out in the experimental spectra, particularly as they are seen in the $\mathsf{MuMax3}$ mode spectra as shown in the Supplement. However, the coupling of these modes is expected to be small, as the higher order quantization reduces the in-phase overlap of the RF field components along $\hat{x}$ and the spin wave dynamic components, so they are less likely to show significant anti-crossings compared to some other laterally quantized modes in $x$ and $y$.

%\section{Conclusion}\label{Conclusion}

We have demonstrated avoided crossings between magnon modes in a micro-platelet of YIG integrated on-chip with a superconducting LC lumped element resonator. 
By transferring the YIG from GGG to a different substrate material, we were able to use a high quality superconducting resonator and achieve anti-crossings for multiple magnon modes quantized in one or more dimensions. For the quasi-uniform mode, the regime of strong-coupling could be reached. The observed anti-crossings agree with analytical modeling and are backed by micro-magnetic simulations. Our work presents another step towards further down-scaling of future on-chip magnon-based microwave devices operating at extremely low power levels as small as 10 fW targeting quantum information science and technologies.

%TC:ignore   
% ^^^ This tells the word counter to ignore the following sections!!!

\section*{Supplementary information}
Additional information regarding details on sample fabrication, experimental procedures, and simulations can be found in the Supplementary Material.
\section*{Author's Contributions}
S.W.K. wrote the manuscript and K. Hu wrote the theoretical sections. P.G. designed and fabricated the superconducting LC resonator circuits, and P.G. and K.He. performed microwave simulations in CST Studio Suite of the LC resonators. F.H. performed FIB milling and placement of the YIG structure onto the LC resonator. S.W.K., P.G., and A. K. performed cryogenic microwave transmission measurements. A. K. and S. W. K. and K. Hu performed $\mathsf{MuMax3}$ simulations and subsequent data analysis. K. Hu and T.P. performed numerical calculations and theory analysis on the experimental data to extract the coupling strength between the magnon modes and LC resonator photon modes. G.S. planned and supervised the experiment and worked on the manuscript. 

G.S. and S.W.K. acknowledge funding by the European Union under Grant Agreement No. 101046630  (Palantiri project). G.S. acknowledges partial funding by the Deutsche Forschungsgemeinschaft in project 490952840 (Harmony) and by the German Research Foundation (DFG) as part of the German Excellence Strategy – EXC3112/1 – 533767171 (Center for Chiral Electronics).  

K. Hu, T.P., and M.E.F. acknowledge support for theoretical simulations from the US Department of Energy, Office of Science, Basic Energy Sciences, under Award No. DE-SC0019250.

\begin{acknowledgments}
The authors thank Rouven Dreyer, Chris Körner, and Hans Hübl for fruitful discussions on interpreting and understanding the experimental and simulation results, and again thank Hans Hübl for supplying designs for the copper sample box used to house the sample in the cryostat. S.W.K. specifically thanks Rouven Dreyer for moral support.
\end{acknowledgments}

\section*{Declarations}
The authors have no conflicting interests to declare.

\section*{Data Availability Statement}
Data openly available in a public repository that issues datasets with DOIs. The data that support the findings of this study are openly available in Zenodo at \href{http://dx.doi.org/10.5281/zenodo.19917945}{http://dx.doi.org/10.5281/zenodo.19917945}.%, reference number [reference number].

\bibliography{YIG_Platelet_Strong_Coupling}

%TC:endignore
\end{document}

\title[Article Title]{SUPPLEMENTARY MATERIAL: Strong coupling between quantized magnon modes in a YIG microstructure and microwaves in a superconducting resonator}

\author[1]{\fnm{Seth W.} \sur{Kurfman}}
\equalcont{These authors contributed equally to this work.}
\author[1]{\fnm{Philipp} \sur{Geyer}}
\equalcont{These authors contributed equally to this work.}
\author[1]{\fnm{Anoop} \sur{Kamalasanan}}

\author[1]{\fnm{Karl} \sur{Heimrich}}

\author[2]{\fnm{Kwangyul} \sur{Hu}}

\author[2]{\fnm{Tharnier O.} \sur{Puel}}

\author[3]{\fnm{Frank} \sur{Heyroth}}

\author[2,4]{\fnm{Michael E.} \sur{Flatté}}

\author*[1,3,5]{\fnm{Georg} \sur{Schmidt}}\email{georg.schmidt@physik.uni-halle.de}

\affil[1]{\orgdiv{Institut für Physik}, \orgname{Martin-Luther-Universität Halle-Wittenberg}, \orgaddress{\city{Halle}, \postcode{06120}, \country{Germany}}}
\affil[2]{\orgdiv{Department of Physics and Astronomy}, \orgname{The University of Iowa}, \orgaddress{\city{Iowa City}, \postcode{52242}, \country{USA}}}
\affil[3]{\orgdiv{Interdisziplinäres Zentrum für Materialwissenschaften}, \orgname{Martin-Luther-Universität Halle-Wittenberg}, \orgaddress{\city{Halle}, \postcode{06120}, \country{Germany}}}

\affil[4]{\orgdiv{Department of Applied Physics and Science Education}, \orgname{Eindhoven University of Technology}, \orgaddress{\city{Eindhoven}, \postcode{5612 AP}, \country{The Netherlands}}}

\affil[5]{\orgname{Halle-Berlin-Regensburg Cluster of Excellence CCE} \country{Germany}}

\maketitle

Note that all available datasets and analysis materials can be found on the Zenodo repository at DOI: \href{http://dx.doi.org/10.5281/zenodo.19917945}{10.5281/zenodo.19917945}.

\section{Methods}\label{Methods}

\subsection{Device Fabrication}

\begin{figure}[h]
    \centering
    \includegraphics[width=0.5\linewidth]{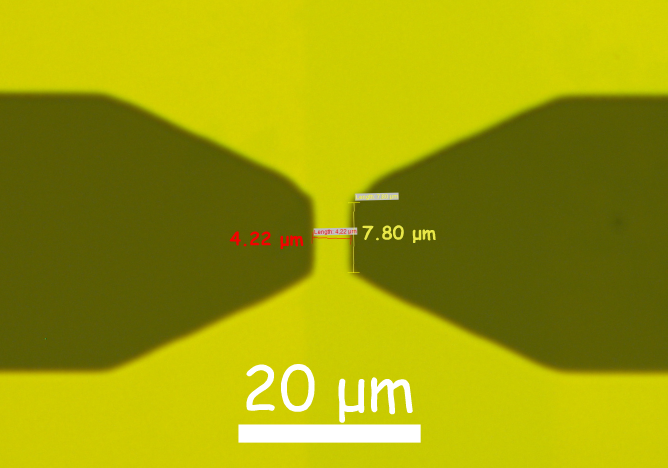}
    \caption{Optical microscope image with dimensions of the inductive element before the YIG platelet was placed. The fact that width of the inductive element is more narrow than the YIG structure strongly influences the selection rules for the excited standing spin wave modes as discussed in the manuscript.}
    \label{supp_fig:inductiveLineBeforeYIG}
\end{figure}

The resonator was fabricated using standard optical lithography processes on a sapphire (0001)-oriented substrate. Prior to patterning and deposition, the substrate was annealed at $1200^{\circ}$ C for 3 hours to reduce surface roughness and correspondingly improve the quality of the deposited resonator material stack. Finally, a tri-layer stack of Nb(5nm)/NbN(30nm)/Nb(7nm) was deposited by magnetron sputtering. An optical image of the inductive line constriction of the bare resonator without YIG is shown in Fig. \ref{supp_fig:inductiveLineBeforeYIG}.

\subsection{FIB Milling of YIG Platelet}\label{Supp_FIBMilling}

A Ga FIB (FEI Versa) was used to mill out a YIG platelet from a commercially purchased (INNOVENT) 790 nm YIG film grown by liquid phase epitaxy (LPE) on a GGG substrate. The lateral dimensions are $\sim (8.6\times 7) \mu m^2$ and the thickness was chosen such that an additional 300 nm of GGG were milled out with the platelet, avoiding radiation damage to the bottom of the platelet. The platelet was then transferred to the inductive line of the LC resonator and fixed in place with Pt welds. Before the milling process, an AlO\textsubscript{x} layer is applied to limit damage to the YIG material from the ion beam.

\subsection{Cryogenic Measurements of Device}\label{Supp_CryoMeasurementStuffs}
 
The samples are loaded into a Scientific Magnetics cryostat with a variable temperature insert (VTI) to achieve temperatures  of $T\sim 1.4$ K  and a 3-D vector magnet capable of applying magnetic fields up to 350 mT along any direction in space. A Tektronix TTR500 Series Vector Network Analyzer (VNA) - with an output power ranging from -50 to 0 dBm - is used to scan the microwave frequency applied to the signal feedline of the LC resonator. The device is mounted within a custom-fabricated copper box, where the internal dimensions are selected to ensure no standing waves are formed in the cavity in the relevant frequency range. The cables in the cryostat included attenuators immediately before and after the copper box to attenuate the DUT input power (before the sample)  and ensure proper thermal anchoring of the copper box to the environment (before and behind the sample).

\subsection{Micromagnetic Simulations} \label{Supp_MuMaxStuffs}

Micromagnetic simulations with $\mathsf{MuMax3}$ are performed to produce magnon spectra and identify the various magnon modes in the YIG platelet \cite{Vansteenkiste2011,Vansteenkiste_2014_AIP-Adv_TheDesignAndVerificationOfMuMax3,Exl2014,Leliaert2017}. The material parameters selected for the simulations of a YIG platelet with dimension $7\times 8.6 $ $\mu$m\textsuperscript{2} with a thickness of $790$ nm broken into $128\times 128\times 16$ cells ($\sim 54.6 \times \sim 67.2\times \sim 49.4$ nm\textsuperscript{3}), a temperature of $T_{sim} = 1.4$ K, and using standard values for YIG saturation magnetization $M_S = 222,929$ A/m ($\mu_0 M_S \sim 280.12$ mT) which is close to the literature value $\mu_0 M = 280$ mT at 2 K (also close to the temperature of our simulation)  \cite{Maier-Flaig_2017_PRB_TemperatureDependentMagneticDampingOfYIGspheres}, exchange constant $A = 3.7\times 10^{-12}$  J/m \cite{Klingler_2015_JPhysDApplPhys_MeasurementsOfTheExchangeStiffnessOfYIGfilmsUsingBroadbandFMRtechniques}, and Gilbert damping $\alpha = 5\times 10^{-4}$ \cite{Klingler_2015_JPhysDApplPhys_MeasurementsOfTheExchangeStiffnessOfYIGfilmsUsingBroadbandFMRtechniques,Maier-Flaig_2017_PRB_TemperatureDependentMagneticDampingOfYIGspheres}. The RF field used to drive the platelet is based on a Biot-Savart law solution to an infinite microstrip waveguide in order to mimic the magnetic fields from the resonator \cite{PGeyer_Bachelorarbeit}. The $\mathsf{MuMax3}$ scripts, data, and analysis codes (Jupyter Notebooks) can be found on the Zenodo repository at DOI: \href{http://dx.doi.org/10.5281/zenodo.19917945}{10.5281/zenodo.19917945}.

\section{Lumped Element Resonator Modes}

\begin{figure}[h]
    \centering
    \includegraphics[width=0.9\linewidth]{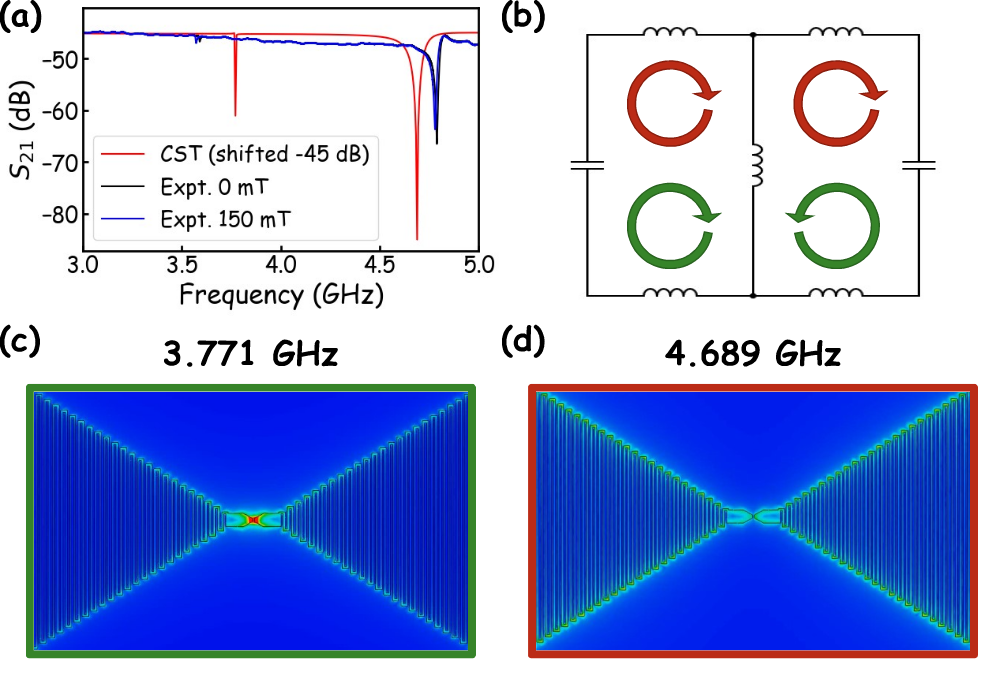}
    \caption{Microwave simulations of the signal line transmission spectrum $S_{21}$ and experimental VNA $S_{21}$ data showing two distinct dips and the corresponding resonator modes. The two dips are denoted "inductive"  and "capacitive," corresponding to the specific mode they represent in the resonator as schematically shown in the effective circuit in panel (b). The red clockwise modes will effectively cancel out the current through the central inductive line and should occur at a higher frequency due to a lower effective inductance ($\omega \propto 1/\sqrt{(LC)}$. The green modes that show the current flow in opposite clockwise and counter-clockwise directions will add in the inductive element. The inductive and capacitive dips in (a) refer to the resonant mode frequency where the mode energy is predominantly located either in the inductive element or in the capacitive elements within the circuit during the oscillation.  This simple circuit model matches well to the inductive and capacitive modes from the CST simulations as seen in panels (c) and (d), respectively. Panels (c) and (d) show the magnitude of the RF field are screenshots from CST Studio. }
    \label{supp_fig:CST_simulations}
\end{figure}

The lumped element LC resonator design was simulated using CST Studio Suite. The symmetry of the resonator leads to one symmetric and one antisymmetric resonator mode with different localization of current density and magnetic field as shown in Fig. \ref{supp_fig:CST_simulations}.

\newpage
\section{Confined-Magnon Mode Maps}

\begin{figure}[h]
    \centering
    \includegraphics[width=0.75\linewidth]{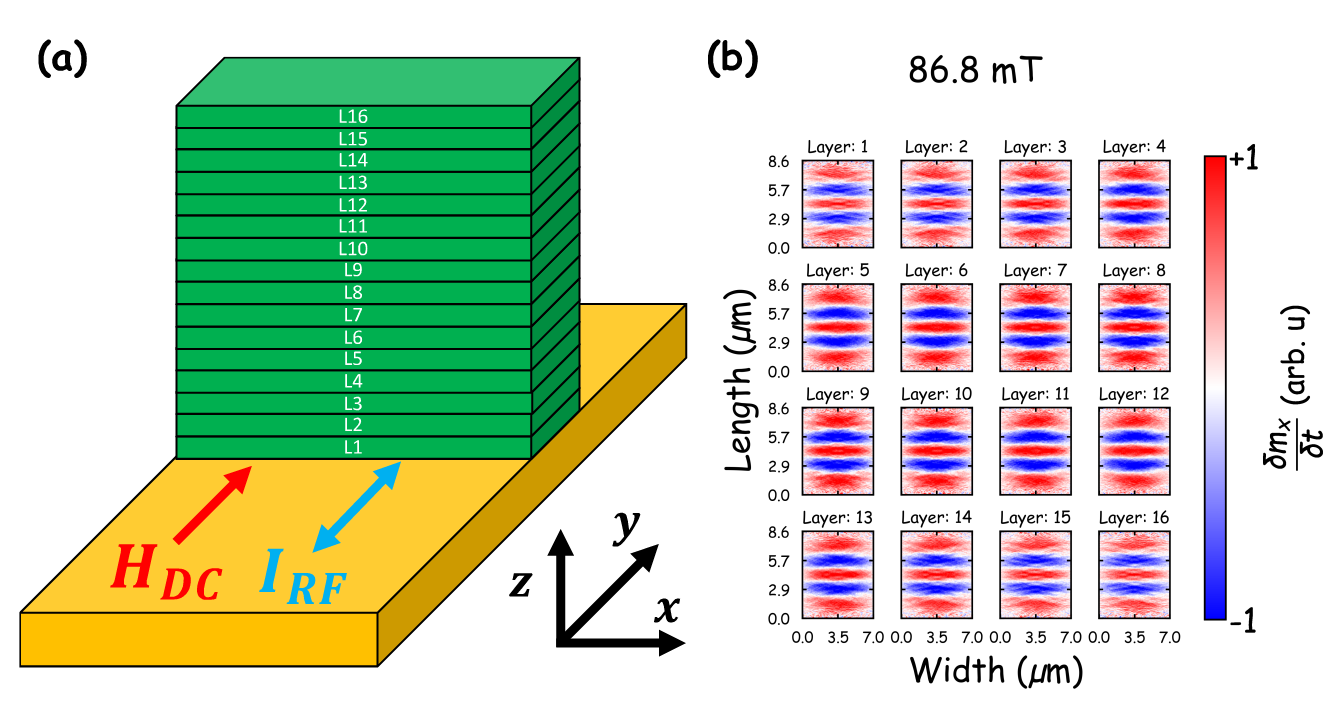}
    \caption{(a) schematic drawing of the YIG platelet with the 16 simulated layers (green) on the inductive line (orange) and the directions of the bias and RF magnetic field. (b) layer resolved mode maps of a BVMSW-like mode with $(m_w,m_l) = (0,5)$ showing that the excitation has only a small change in intensity and constant phase throughout the layer thickness.}
    \label{supp_fig:modeMaps_0degree_layers}
\end{figure}

The magnon mode maps shown in Fig. 3 of the manuscript were showing the spin dynamics in a cross-section of the platelet in the middle layer (specifically, Layer 8 for comparison to Fig. \ref{supp_fig:modeMaps_0degree_layers}). For one of the modes, specifically $(m_w,m_l) = (0,5)$, Fig. \ref{supp_fig:modeMaps_0degree_layers} shows that the dynamics are more or less identical throughout the film thickness. This was also confirmed for all lateral quantized BVMSW-like modes.

\newpage

\section{Fine Measurements of the Main Anti-Crossing}
\begin{figure}[h]
    \centering
    \includegraphics[width=1\linewidth]{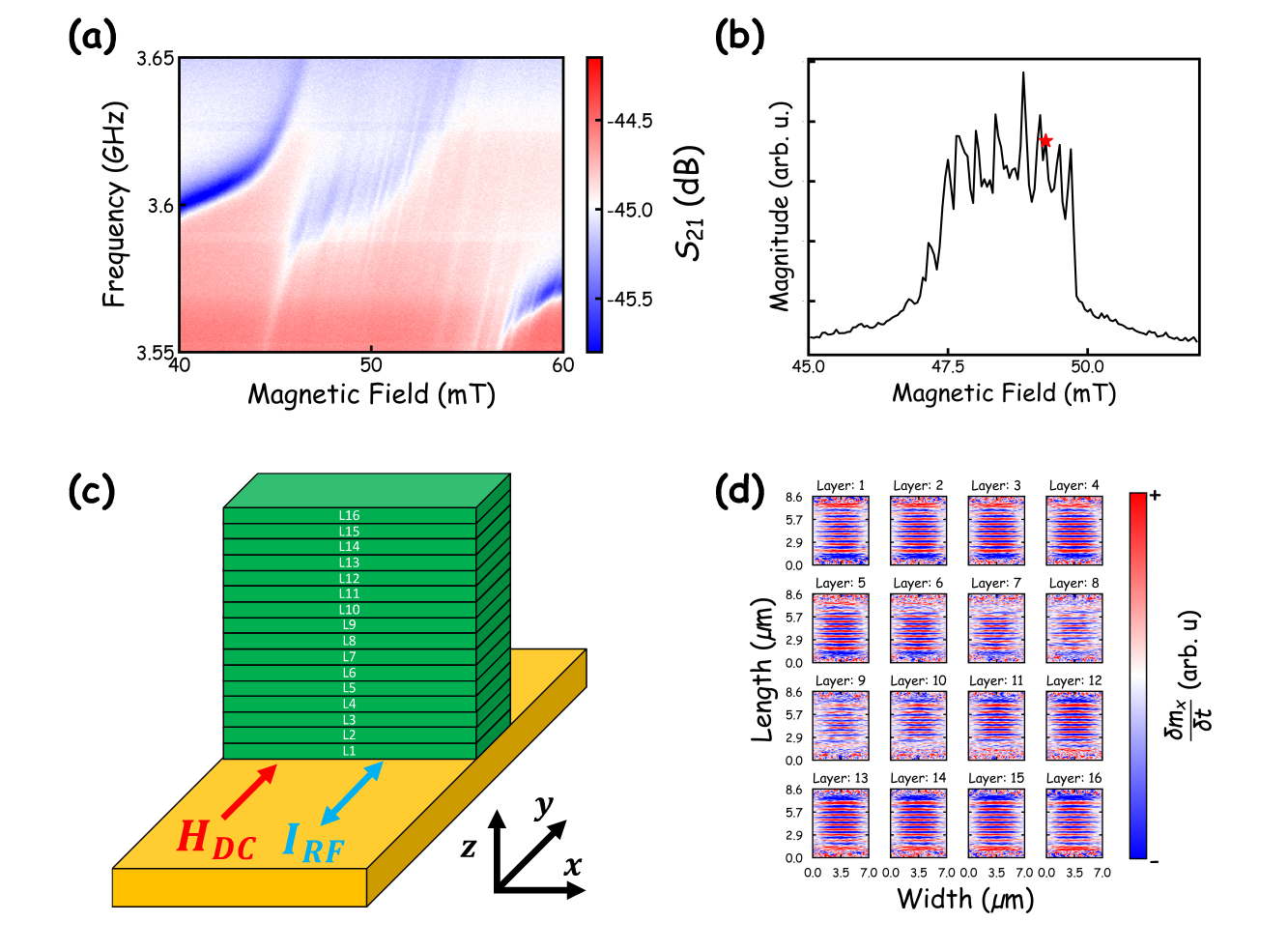}
    \caption{(a) Zoom-in and fine scan of the large anti-crossing region shown in Fig. 2 of the main text. Note the range of the colorbar is the same as in the main text, but the center value has been shifted by +0.275 dB to accent the anti-crossings in the middle which likely correspond to higher-order magnon modes around the main peak. (b) $\mathsf{MuMax3}$ simulation-derived spectra for the broad (0,0) mode which causes the main anti-crossing. The star indicates the position at which the mode maps below were evaluated. (c) Schematic drawing of the YIG platelet with the 16 simulated layers (green) on the inductive line (orange) and the directions of the bias and RF magnetic field. (d) Layer-resolved mode maps of a PSSW-like mode at the magnetic field marked by a red star in (b) showing that the excitation has maxima at the top and bottom of the film but exhibits a 180 $\degree$ phase-change between top and bottom.}
    \label{supp_fig:Experiment_ZoomIn}
\end{figure}

As mentioned in the main text, the primary anti-crossing between $\sim45-55$ mT shows in fact a number of additional modes seen in Fig. \ref{supp_fig:Experiment_ZoomIn}. Similar modes have been seen in other work in the literature between the primary anti-crossing \cite{Xu_2024_AdvScience_StrongPhotonMagnonCouplingUsingALithographicallyDefinedOrganicFerrimagnet,Zhang_2016_JAP_SuperstrongCouplingofThinFilmMagnetostaticWavesWithMicrowaveCavity}. Our $\mathsf{MuMax3}$ simulation indicate that these may be PSSW modes with quantization along the thickness of the platelet, which, due to their asymmetry, are only weakly excited. 

In Fig. \ref{supp_fig:Experiment_ZoomIn} we show a $\mathsf{MuMax3}$ simulation for the broad (0,0) mode and layer resolved mode maps for one selected PSSW mode from the main anti-crossing caused by the (0,0) mode. The mode map confirms the antisymmetric perpendicular quantization with maximum amplitude at top and bottom and a phase shift of $180\degree$ through the film.

%\begin{figure}[h]
%    \centering
%    \includegraphics[width=0.75\linewidth]{Figures/v5_2026-18-03/SuppFig_MuMaxStructure.png}
%    \caption{Simple schematic showing the excitation waveguide (gold) used in the $\mathsf{MuMax3}$ simulations to drive the RF field via an RF current $I_{RF}$ for the YIG platelet (green). The platelet is broken up into 16 layers labeled $L\#$. This is used as a guide for the following Figures \ref{}-\ref{}, where the laterally-resolved quantized modes from the simulation are plotted for each layer of the structure at various magnetic fields. These mode maps align well qualitatively with the predicted 3D-quantized mode mixing given by the analytical calculations in the Main text.}
%    \label{supp_fig:schematicOfMuMaxSetup}
%\end{figure}

%\subsubsection{$\theta = 0\degree$ Orientation}

\newpage

\section{Power-Dependence of Measurements}

\begin{figure}[h]
    \centering
    \includegraphics[width=0.75\linewidth]{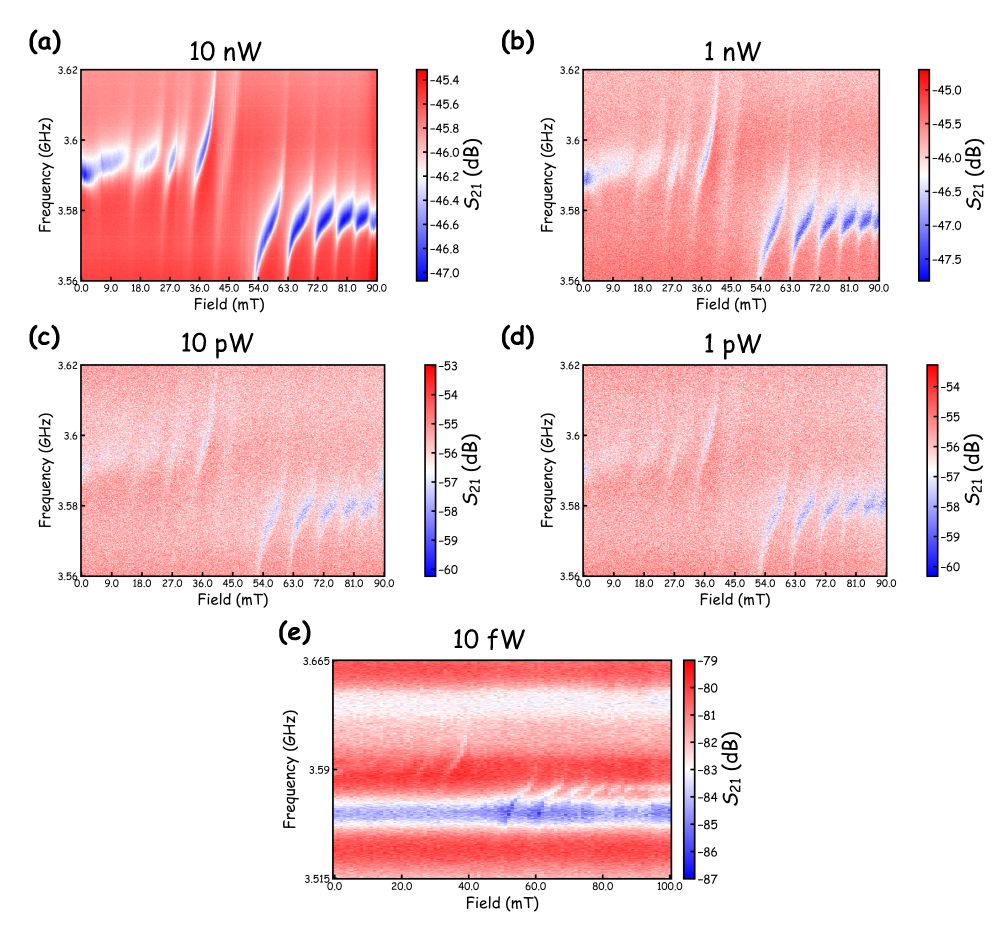}
    \caption{Experimental heat maps performed at a variety of DUT input powers, namely (a) 10 nW, (b) 1 nW, (c) 10 pW, (d) 1 pW. Note these measurements use a mix of attenuators to reduce the power into the DUT, and that there was no need for additional amplification in order to measure these anti-crossing signatures. In panel (e), the DUT input power is 10 fW, where a 40 dB low noise amplifier (LNA) was used at the output side of the DUT in order to increase the signal above the noise floor of the VNA.}
    \label{supp_fig:powerDependence}
\end{figure}

As described in the manuscript, the anti-crossings are detectable at very low excitation power levels. In Fig. \ref{supp_fig:powerDependence}, we show the measurements for a variety of low input powers to the DUT. Notably, the device shows indications of the anti-crossing features for $\theta = 0\degree$ even down to DUT powers of -110 dBm (10 fW) seen in Fig. \ref{supp_fig:powerDependence}(e).

\newpage

\bibliography{YIG_Platelet_Strong_Coupling}